\documentclass[twocolumn]{aastex63}
\usepackage{xfrac,graphicx,amsmath,amssymb,balance}

\newcommand{\heii}{He{\sc ii}}
\newcommand{\civ}{C{\sc iv}}
\newcommand{\ovi}{O{\sc vi}}
\newcommand{\oiii}{O[{\sc iii}]}

\received{October 1, 2024}
\revised{October 1, 2024}
\accepted{October 1, 2024}
\submitjournal{ApJ}

\shorttitle{A hot pulsating pre-white dwarf from OGLE}
\shortauthors{Pietrukowicz et al.}

\begin{document}

\title{An Extremely Hot Pulsating Pre-White Dwarf from OGLE}

\correspondingauthor{Pawe{\l} Pietrukowicz}
\email{pietruk@astrouw.edu.pl}

\author[0000-0002-2339-5899]{Pawe{\l} Pietrukowicz}
\affiliation{Astronomical Observatory, University of Warsaw, Al. Ujazdowskie 4, 00-478 Warszawa, Poland}

\author[0000-0002-6428-2276]{Klaus Werner}
\affiliation{Institut f\"ur Astronomie und Astrophysik, Kepler Center for Astro and Particle Physics, Eberhard Karls Universit\"at, Sand 1, 72076 T\"ubingen, Germany}

\author[0000-0002-8911-6581]{Mateusz J. Mr\'oz}
\affiliation{Astronomical Observatory, University of Warsaw, Al. Ujazdowskie 4, 00-478 Warszawa, Poland}

\author[0000-0002-3218-2684]{Milena Ratajczak}
\affiliation{Astronomical Observatory, University of Warsaw, Al. Ujazdowskie 4, 00-478 Warszawa, Poland}


\begin{abstract}
We show that the blue 18.3-minute variable object discovered in the Galactic disk by the OGLE-III survey and named OGLE-GD-WD-0001 is a pulsating pre-white dwarf of PG 1159 spectral type. With an effective temperature of about 160,000~K it is among the hottest known pulsators being located close to the blue edge of the GW~Virginis instability strip. The long-term OGLE observations indicate that the object has a positive period change rate of about $5\times10^{-10}$ s~s$^{-1}$ and thus already contracts. There are no traces of a planetary nebula around this star.
\end{abstract}


\section{Introduction}\label{sec:intro}

Pulsations are observed in stars of all luminosity classes, including white dwarfs (WDs) and objects being close to this final evolutionary stage for low-mass stars \citep{2022ARA&A..60...31K}. Among WDs, three main types of pulsating variables are distinguished: ZZ~Ceti, V777 Herculis, and GW~Virginis stars \citep{2008PASP..120.1043F}. They represent the following three spectral classes, respectively: hydrogen-rich DAV stars with effective temperatures $10,400 < T_{\rm eff} < 12,400$~K and surface gravities $7.5 < \log g < 9.1$ (in the units of cm~s$^{-2}$), helium-rich DBV stars with $22,400 < T_{\rm eff} < 32,000$~K and $7.5 < \log g < 8.3$, and helium/carbon/oxygen-rich DOV stars with $80,000 < T_{\rm eff} < 180,000$~K and $5.5 < \log g < 7.5$ \citep{2019A&ARv..27....7C}. Stars with $T_{\rm eff} > 100,000$~K are classified as white dwarf precursors (pre-WDs) and are often surrounded by a nebula. It is assumed that pulsations in WDs and pre-WDs are generated by global nonradial gravity modes. The brightness variations are usually irregular, of millimagnitudes or lower, very rarely exceed 0.1 mag. The observed pulsation periods range from 1.6--23.3~min in the ZZ~Cet stars, 2.0--18~min in the V777 Her stars, and 5--100~min in the GW~Vir stars. WDs and pre-WDs are multi-periodic pulsators, but many modes, due to their low amplitudes, can be detected only from space \citep[e.g.,][]{2020A&A...638A..82B,2022A&A...668A.161C,2024A&A...686A.140C}. Period change rates in GW~Vir stars are of the order of $10^{-10}$--$10^{-12}$~s~s$^{-1}$, two orders of magnitude higher than in V777 Her stars, and four--five orders of magnitude higher than in ZZ~Cet stars \citep{2019A&ARv..27....7C}. The GW~Vir instability domain is partly populated by PG 1159 = DOV stars, being likely the progenitors of hydrogen-deficient WDs. The other group of pre-WDs is formed by central stars of planetary nebulae (PNe) with carbon-rich Wolf–Rayet spectra. According to the census published in \cite{2023ApJS..269...32S}, only 24 out of 67 known PG 1159 stars are confirmed to pulsate. Twenty-nine PG 1159 stars are not surrounded by bright planetary nebulae. The detection and proper classification of short-period objects often require both photometric and spectroscopic data.

\section{The OGLE survey}\label{sec:data}

The Optical Gravitational Lensing Experiment (OGLE) is a long-term photometric survey of the Milky Way stripe and Magellanic Clouds conducted from Las Campanas Observatory, Chile, using the 1.3-m Warsaw telescope. One of the main scientific goals is the search for microlensing events toward the Galactic bulge to characterise dark and low-luminosity objects, such as black holes \citep{2024Natur.632..749M} and extrasolar planets \citep[e.g.,][]{2017Natur.548..183M}. Since 2010 the survey is in its fourth phase \citep[OGLE-IV,][]{2015AcA....65....1U}. Thanks to a wide-field (1.4 deg$^2$) 32-detector mosaic camera, OGLE-IV monitors around 2 billion stars over an area of about 3600 deg$^2$. The previous phase, OGLE-III, was conducted in years 2001--2009 and covered about 170 deg$^2$ of the sky. The OGLE-III camera was a mosaic of eight detectors with a total field of view of 0.35 deg$^2$. OGLE images are taken mainly in the $I$-band, while additional $V$-band data secure color information. The data are reduced with the help of Difference Image Analysis \citep[DIA,][]{2000AcA....50..421W}, especially developed for dense stellar fields.

During the third phase of the OGLE project, a 7.12-deg$^2$ area of the Galactic disk (GD) in the direction tangent to the Carina Arm (galactic longitudes between $+288\degr$ and $+308\degr$) was surveyed with the prime aim to detect transiting extrasolar planets \citep{2002AcA....52..317U,2004AcA....54..313U}. The observations were conducted with a cadence of about 16 min. Twenty-one selected GD fields were monitored alternately for several hours every clear night. Searches for variable stars in the disk area led to the discovery of thousands of eclipsing binary stars \citep{2013AcA....63..115P} and hundreds of pulsating stars of various types \citep{2013AcA....63..379P}, including the first Blue Large-Amplitude Pulsator \citep[BLAP,][]{2017NatAs...1E.166P}. One of the detected variable stars was a blue object (with a color of $V-I=-0.23$ mag) exhibiting sinusoidal light variations with an amplitude of about 0.01 mag in the $I$-band at a period of about 18.345 min or 1100.7~s. The object was tentatively classified as a pulsating white dwarf and named OGLE-GD-WD-0001 \citep[][alternative name UCAC4 141-049067, equatorial coordinates RA(2000.0) = 10:39:58.06, DEC(2000.0) = $-$61:57:32.1, galactic coordinates $l=+288\fdg1312$, $b=-2\fdg9257$]{2013AcA....63..379P}.

However, the relatively high brightness ($V=14.29$, $I=14.52$ mag) and the parallax of $1.30\pm0.03$ mas from Gaia DR3 \citep{2021A&A...649A...1G} questioned the proposed class of this stars as a normal WD. The object seemed to be hotter and more luminous. It was proposed to be a hot subdwarf candidate \citep{2019A&A...621A..38G} based on Gaia DR2 \citep{2018A&A...616A...1G}.

\section{Photometric data}\label{sec:phot}

\begin{table*}[]
\centering \caption{Consize log of photometric observations}
{\fontsize{7}{8} \selectfont
\begin{tabular}{ccccccccc}
\hline
Season  &  Period of observations  &      BJD$-$2450000.0     & $N_{\rm nights}$ & $N_{\rm exp}$ & $N_{\rm texp=120~s}$ & $N_{\rm texp=180~s}$ & $r=$ & Median of $r$\\
        &                          &                          &  w/o gaps        &               &                      &                      & $t_{\rm exp}$/duty cycle     & \\
\hline
2005    & 2005-02-01~--~2005-06-23 & 3402.87333~--~3544.58296 & 142/104          &  1323         &  1321                &      2               & 0.035~--~0.186 & 0.123 \\
2006    & 2006-01-13~--~2006-02-16 & 3748.85706~--~3782.83887 &  34/ 31          &   162         &   162                &      0               & 0.123~--~0.248 & 0.148 \\
2007    & 2006-10-30~--~2007-07-28 & 4038.79072~--~4309.53905 & 271/116          &  1558         &  1226                &    332               & 0.051~--~0.621 & 0.129 \\
2008    & 2007-12-27~--~2008-06-16 & 4461.81492~--~4633.50806 & 172/166          &   782         &   782                &      0               & 0.062~--~0.260 & 0.143 \\
\hline
Overall &                          &                          & 417              &  3825         &  3491                &    334               & 0.035~--~0.621 & 0.133 \\
\hline
\label{tab:log}
\end{tabular}}
\end{table*}

\begin{figure}
\centering
\includegraphics[width=0.46\textwidth]{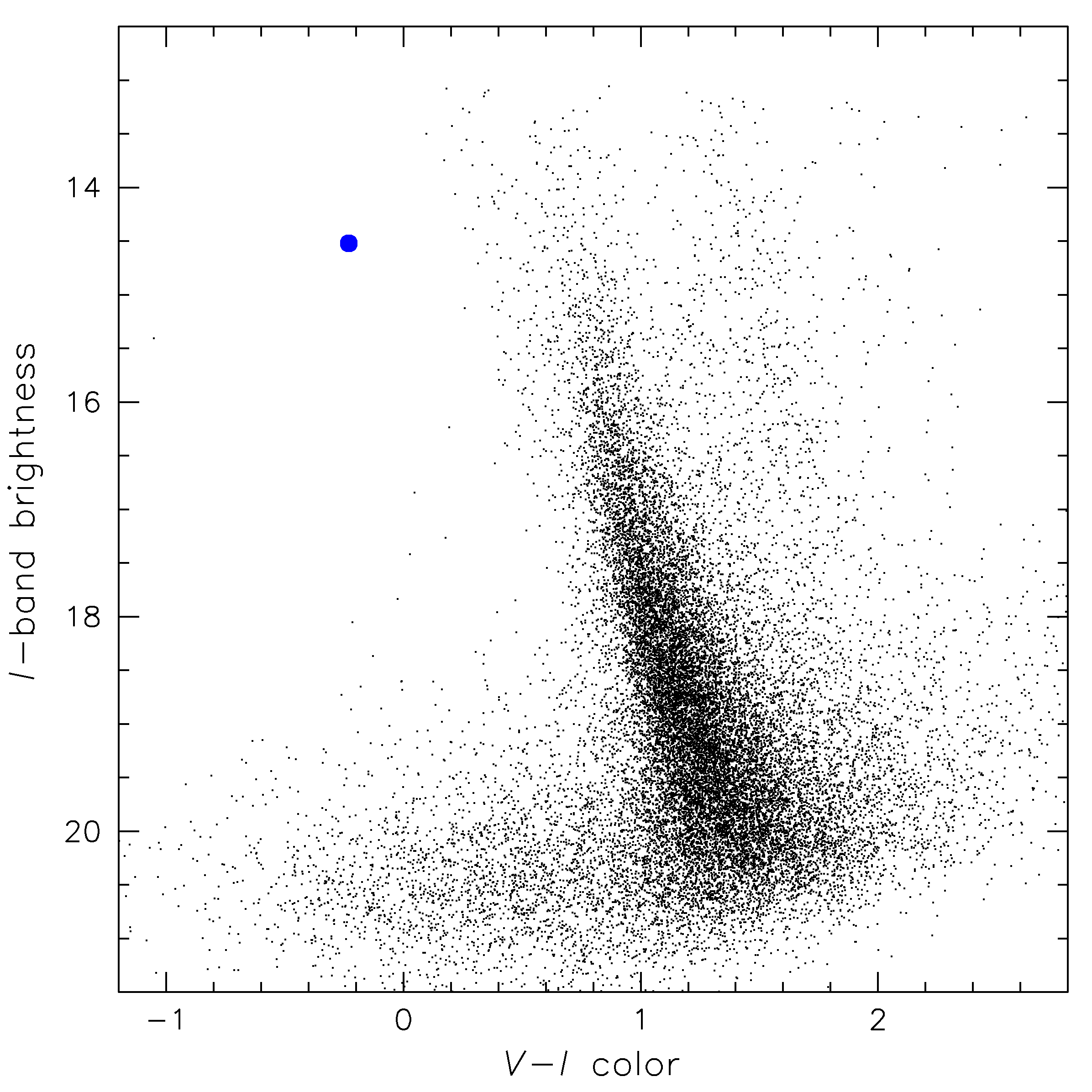}
\caption{Color-magnitude diagram with the location of the blue object OGLE-GD-WD-0001.}
\label{fig:cmd}
\end{figure}

\begin{figure}
\centering
\includegraphics[width=0.46\textwidth]{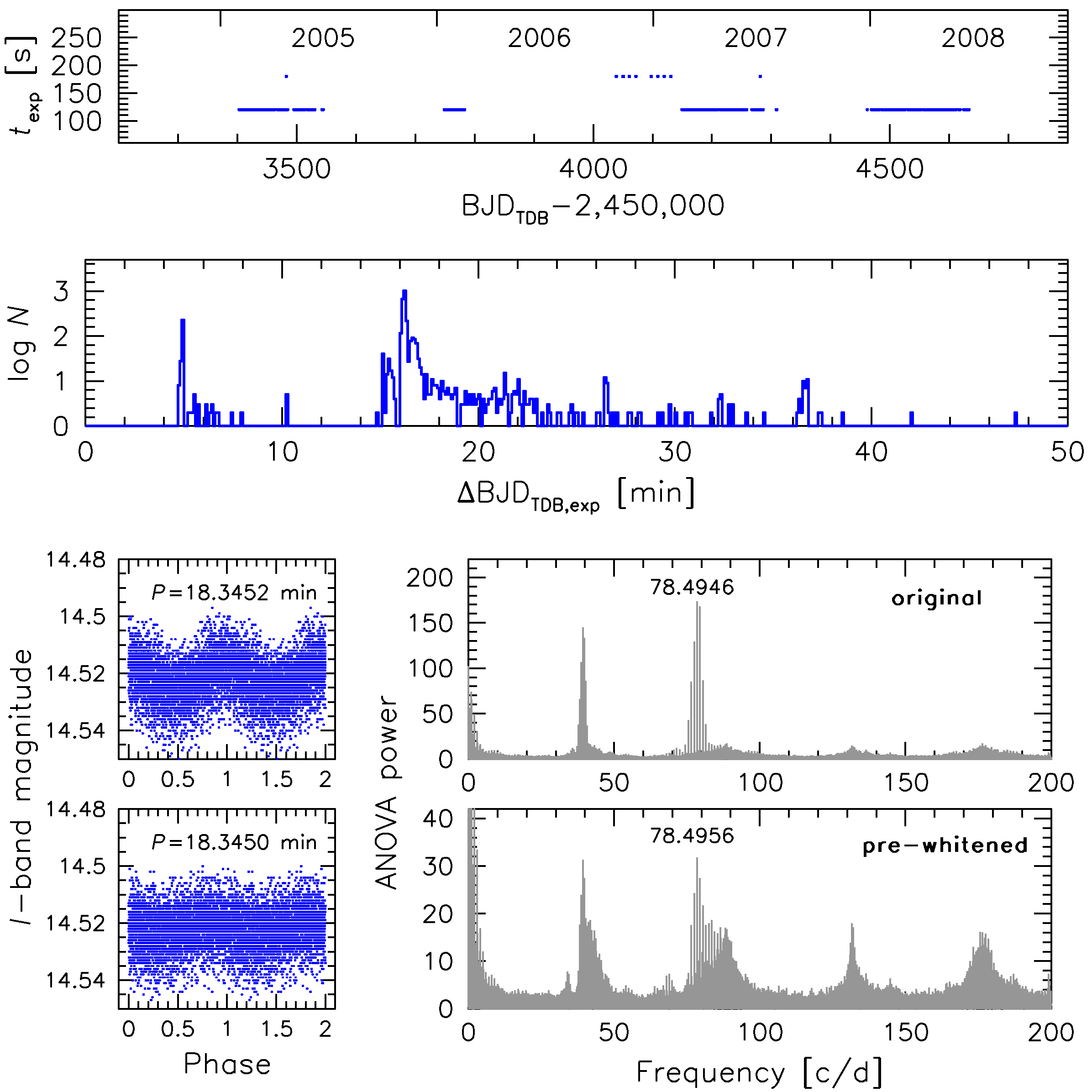}
\caption{OGLE-III $I$-band time-series data for object OGLE-GD-WD-0001 and results of the frequency analysis: distribution of the observations in time (top panel), cadence of the observations shown as a histogram of time differences between two subsequent exposures (second panel from the top), phase-folded original and pre-whitened light curves together with the power spectra (lower panels). The frequency is given in cycles per day.}
\label{fig:power}
\end{figure}

The photometric data for star OGLE-GD-WD-0001 are available at the OGLE Internet Archive\footnote{\url{https://ftp.astrouw.edu.pl/ogle/}}. The star is located in two overlapping OGLE-III fields, CAR109 and CAR115, which were observed in the $I$-band more than, in total, 3800 times over 417 nights in years 2005--2008. Most of the images (about 91 per cent) were taken with the exposure time $t_{\rm exp}=120$~s, while the remaining images with $t_{\rm exp}=180$~s. Six additional measurements were obtained in the $V$-band with the exposure time of 200~s. Fig.~\ref{fig:cmd} illustrates the position of the star in a color-magnitude diagram constructed based on the photometric maps in \cite{2010AcA....60..295S} for subfield CAR115.5. In Tab.~\ref{tab:log}, we provide a log of observations consize to seasons. In the table, we introduce the $r$ value which is the ratio of the exposure time to the duty cycle calculated on nightly basis. The distribution of the $I$-band observations in time is shown in the upper panel of Fig.~\ref{fig:power}. In the figure, we also show cadence of the observations as a histogram of differences between the middle moments of two subsequent exposures. The main peak is around 16.2~min (88.9 cycles per day), while the second highest peak is around 4.0~min (360 c/d). This reflects the observing strategy in which either four selected Galactic disk fields were monitored alternately or one of the field was monitored continuously. For the analysis we combined the time-series photometry from the two fields by adding a brightness offset to the light curve from field CAR109 to the more rich data from field CAR115. We converted moments of the brightness measurements from heliocentric Julian date (HJD) to more accurate barycentric Julian date (BJD$_{\rm TDB}$). The time-series data were analyzed with the TATRY code \citep{1996ApJ...460L.107S}, which employs periodic orthogonal polynomials to fit the data and the analysis of variance statistics to evaluate the quality of the fit. We removed several evident outlying points from the phase-folded light curve and corrected the period. A single sinusoid was fit to the data. We found the variability period of $P=0.0127397220(28)$~d = 18.345200(4) min = 1100.7120(2)~s and $I$-band amplitude of 0.0090(1)~mag. The detected signal is strong and independent to the cadence of observations. The period uncertainty was calculated based on the width of the frequency peak in the power spectrum. The original light curve together with the power spectrum are shown also in Fig.~\ref{fig:power}. In the power spectrum, the highest peak (around frequency $f=1/P=78.4946$ c/d) corresponds to the true variability period, while the second highest peak is its alias ($f/2=39.2474$ c/d). The latter period would result in an unrealistic shape of the phased light curve (as a squeezed sinusoid with two maxima over the cycle). The remaining low and wide peaks in the power spectrum stem from the cadence of observations and their multiplicities. The detected period is either about 9.1 times longer (for 91 per cent of the observations) or 6.1 times longer than the applied exposure times, which slightly reduces the true amplitude. The reduction factor can be calculated as ${\rm sinc}(\pi P/t_{\exp})$ and its value is 0.981 for $t_{\rm exp}=120$~s and 0.957 for $t_{\rm exp}=180$~s. After pre-whitening the original light curve, we found an additional signal with a period of 0.0127395595(68)~d = 18.344966(10) min = 1100.6979(6)~s and $I$-band amplitude of 0.0038(1)~mag (see bottom part of Fig.~\ref{fig:power}). This period is very close to the first (original) value.

We note that the star is also observed in OGLE-IV, but with a much lower frequency (typically several times per month).

\section{Spectroscopic data}\label{sec:spec}

\begin{figure*}
\centering
\includegraphics[width=0.90\textwidth]{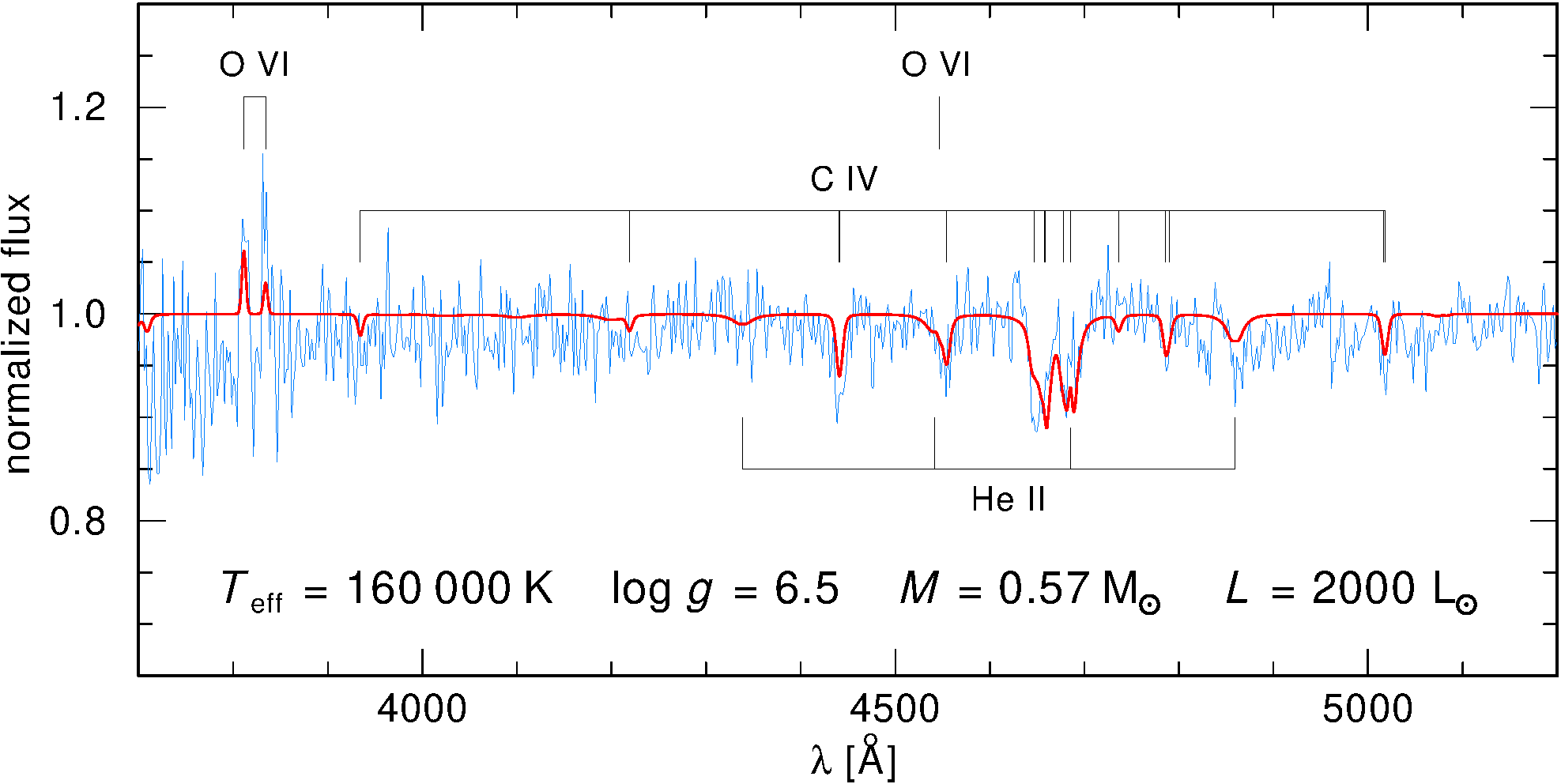}
\caption{Normalized spectrum of OGLE-GD-WD-0001 together with the best fit.}
\label{fig:spec}
\end{figure*}

The spectroscopic observations of OGLE-GD-WD-0001 were obtained with the EFOSC2 spectrograph \citep{1984Msngr..38....9B} on the 3.58-m European Southern Observatory (ESO) New Technology Telescope (NTT) located at La Silla Observatory, Chile, under ESO programme 113.C-0250(A). We used grism \#7 covering spectral wavelengths from 3270--5240~{\AA} at a slit width of $1\farcs0$. The observations were conducted under clear sky but not good seeing conditions (around $1\farcs35$). Two 190-s exposures were taken to increase the signal to noise and to reject possible cosmic rays. The measured resolution of the spectrum is 5~{\AA} at 4000~{\AA}. The data were reduced using the standard EFOSC2 Pipeline (version 2.3.9), which performed bias and sky subtraction, flat-fielding, wavelength calibration, and spectrum extraction. In Fig.~\ref{fig:spec}, we present the obtained normalized spectrum after coadding the two single exposures.

\section{Properties of OGLE-GD-WD-0001}\label{sec:param}

The observed brightness variations in OGLE-GD-WD-0001 are typical for pulsating PG1159 stars. Periods in these stars are in the range of 5--100 min, while amplitudes may reach 0.15 mag \citep{2019A&ARv..27....7C}. The two very closely located peaks in the power spectrum of OGLE-GD-WD-0001 are very likely a result of a long-term period change. However, we cannot rule out a possibility that the presence of the additional peak stems from irregular phase and/or amplitude changes. By dividing the OGLE-III data into two sections of time, seasons 2005--2006 and 2007--2008, we found a period change rate of $dP/dt=(+4.9\pm0.3) \times 10^{-10}$~s~s$^{-1}$. This value is in agreement with rates measured in other PG1159 stars \citep[e.g.,][]{2019A&ARv..27....7C,2021A&A...645A.117C} and indicate that our object already evolves toward lower effective temperatures and contracts.

The spectrum of OGLE-GD-WD-0001 exhibits a broad absorption-line trough in the wavelength range 4650--4700\,{\AA}, which is a unique signature of PG1159 stars \citep[see e.g.,][]{2006PASP..118..183W}. It is made up by several lines from {\civ} and {\heii} 4686\,{\AA}. Other prominent {\civ} lines are detectable at 4441, 4554, and 5017\,{\AA}. There is a strong {\ovi} 3811/3834~{\AA} doublet in emission, indicating a very high effective temperature of the star.

For the spectral analysis we used a grid of non-local thermodynamic equilibrium (NLTE) models that was computed with the T\"ubingen Model-Atmosphere Package (TMAP\footnote{\url{http://astro.uni-tuebingen.de/~TMAP}}). The line-blanketed models assume plane-parallel geometry and are in radiative and hydrostatic equilibrium \citep{1999JCoAM.109...65W,2003ASPC..288...31W,2012ascl.soft12015W}. The models are of the type introduced in detail by \cite{2014A&A...564A..53W} and were tailored to investigate the optical spectra of hot hydrogen-deficient PG1159 stars. In essence, they consist of the main atmospheric constituents, namely helium, carbon, and oxygen.

Our best fitting model indicates that the atmosphere is composed of helium and carbon in equal proportions, with a 0.005 mass fraction of oxygen. We obtained an effective temperature of $T_{\rm eff}=160,000\pm30,000$~K ($\log~T_{\rm eff} \approx 5.2$) and a surface gravity of $\log~g=6.5\pm0.5$. From evolutionary tracks \citep{2006A&A...454..845M}, we found the mass of the star $M=0.57$~M$_{\odot}$ and the luminosity of about $L=2000$~L$_{\odot}$ or $\log(L/{\rm L}_{\odot})\approx3.3$.

\begin{figure*}
\centering
\includegraphics[width=0.70\textwidth]{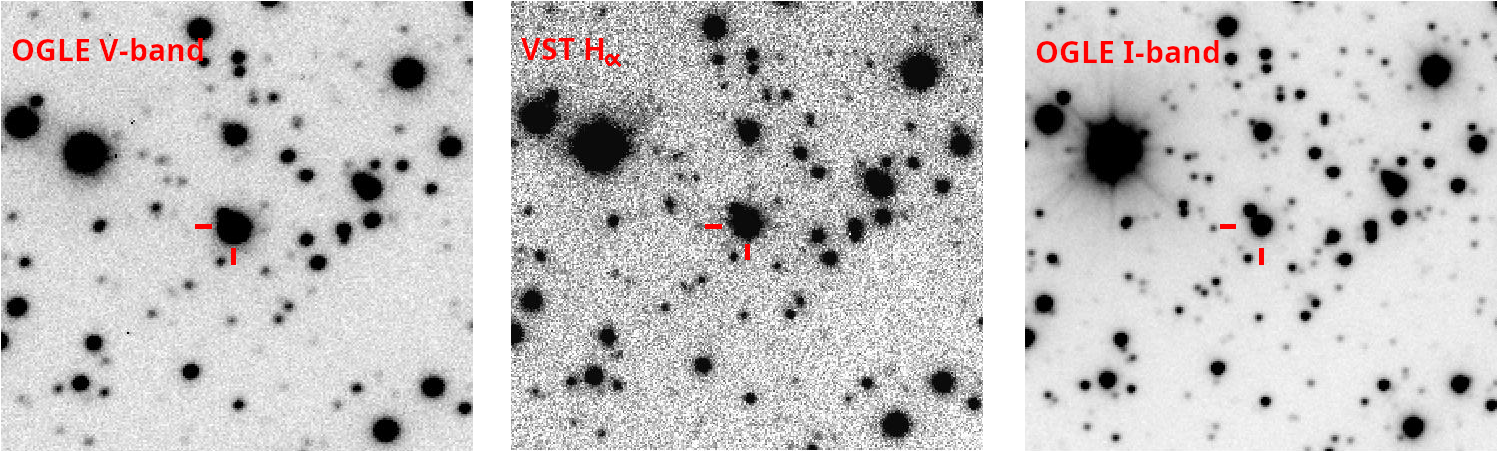}
\caption{Images with star OGLE-GD-WD-0001 (in their centers) in three different bands: $V$ (left panel) and $I$ (right panel) from the OGLE survey and $H\alpha$ (middle panel) from the VPHAS+ survey \citep[conducted on the ESO VLT Survey Telescope,][]{2014MNRAS.440.2036D}. Each image is $1\farcm0$ on the side. North is up and east is to the left. The OGLE images are cropped templates for image subtraction that were stacked from several single frames.}
\label{fig:charts}
\end{figure*}

We report that there are no traces of a nebula around OGLE-GD-WD-0001. In Fig.~\ref{fig:charts}, we present $1\farcm0 \times 1\farcm0 $ images with the object in the center in three available optical bands: through broadband $V$- and $I$-filters from the OGLE 1.3-m Warsaw telescope and the narrowband $H\alpha$ filter from the 2.6-m ESO VLT Survey Telescope \citep[collected in the course of the VPHAS+ survey,][]{2014MNRAS.440.2036D}. The OGLE images are templates stacked from several high-quality individual frames and used in the data reduction process. Any fine structures would emerge from the background. The obtained spectrum exhibits no emission lines characteristic for PNe, such as {\oiii} 4959/5007\,{\AA} doublet.

\section{Conclusions}\label{sec:conc}

We showed that the blue short-period variable object OGLE-GD-WD-0001 is a very hot and luminous pre-WD of PG 1159 spectral type. The obtained physical and atmospheric parameters, $T_{\rm eff} = 160,000\pm30,000$\,K, $\log g=6.5\pm0.5$, the luminosity of 2000 L$_{\odot}$, and the period of about 18.345 min clearly indicate that this is a rare pulsating star of GW~Vir type. With this effective temperature, OGLE-GD-WD-0001 is among the hottest GW~Vir pulsators being close to the blue edge of the GW~Vir instability strip \citep[see Fig.\,6 in][]{2023ApJS..269...32S}. It is very similar to the GW~Vir-type variable object SALT J172411.7$-$632147 of the same $T_{\rm eff}$ and gravity, and pulsation periods between about 11--17 min \citep{2023MNRAS.519.2321J}. According to the list provided in \cite{2023ApJS..269...32S}, and taking into account the recent discoveries presented in \cite{2023MNRAS.519.2321J}, there are nine similar or hotter than OGLE-GD-WD-0001 pulsating PG 1159 stars identified, four of which are not surrounded by a nebula. Nevertheless, a higher resolution spectrum would allow for a more accurate determination of the physical parameters of our object.

\acknowledgements
We thank Nicole Reindl for discussion on the obtained spectrum. This work is based on observations collected at the European Organisation for Astronomical Research in the Southern Hemisphere under ESO program 113.C-0250(A). We used data from the European Space Agency (ESA) mission Gaia, processed by the Gaia Data Processing and Analysis Consortium (DPAC). Funding for the DPAC has been provided by national institutions, in particular the institutions participating in the Gaia Multilateral Agreement.

\end{document}